# A Stochastic Process Approach of the Drake Equation Parameters

Nicolas Glade[1], Pascal Ballet[2] and Olivier Bastien[3, *]

[1] *Joseph Fourier University, AGeing, Imagery and Modeling (AGIM) Laboratory, CNRS FRE3405, Faculty of Medicine of Grenoble, 38700 La Tronche, France*

[2] *European University of Brittany (UEB) - University of Brest, Complex Systems and Computer Science Laboratory (LISyC) - EA3883, 20 avenue LeGorgeu, 29238, Brest Cedex*

[3] *Laboratoire de Physiologie Cellulaire Végétale. UMR 5168 CNRS-CEA-INRA-Université Joseph Fourier, CEA Grenoble, 17 rue des Martyrs, 38054, Grenoble Cedex 09, France*

(*) Corresponding author:

Olivier Bastien, olivier.bastien@cea.fr.

tel. +33 (0)4 38 78 38 55.

fax +33 (0)4 38 78 50 91

**Abstract.** The number $N$ of detectable (i.e. communicating) extraterrestrial civilizations in the Milky Way galaxy is usually done by using the Drake equation. This equation was established in 1961 by Frank Drake and was the first step to quantifying the SETI field. Practically, this equation is rather a simple algebraic expression and its simplistic nature leaves it open to frequent re-expression An additional problem of the Drake equation is the time-independence of its terms, which for example excludes the effects of the physico-chemical history of the galaxy. Recently, it has been demonstrated that the main shortcoming of the Drake equation is its lack of temporal structure, i.e., it fails to take into account various evolutionary processes. In particular, the Drake equation doesn't provides any error estimation about the measured quantity. Here, we propose a first treatment of these evolutionary aspects by constructing a simple stochastic process which will be able to provide both a temporal structure to the Drake equation (i.e. introduce time in the Drake formula in order to obtain something like $N(t)$ ) and a first standard error measure.

## 1. Introduction

The number of detectable (i.e. communicating) extraterrestrial civilizations in the Milky Way galaxy is usually done by using the Drake equation (Burchell, 2006). This equation was established in 1961 by Frank Drake and was the first step to quantifying the Search for ExtraTerrestrial Intelligence (henceforth SETI) field (Drake, 1965). This formula is broadly used in the fields of exobiology and the SETI. Practically, this equation is rather a simple algebraic expression and its simplistic nature leaves it open to frequent re-expression (Walters *et al.*, 1980; Shermer, 2002; Burchell, 2006; Forgan, 2009). While keeping in mind that other equivalent forms exist, we investigate the following form:

$$N^* = R^* f_p n_e f_l f_i f_c L \qquad (1)$$

In this expression, the symbols have the following meanings: $N$= the number of Galactic civilizations who can communicate with Earth; $R^*$ = the average rate of star formation per year in our galaxy; $f_p$ = the fraction of stars that host planetary systems; $n_e$ = the number of planets in each system that are potentially habitable; $f_l$ = the fraction of habitable planets where life originates and becomes complex; $f_i$ = the fraction of life-bearing planets that bear intelligence; $f_c$ = the fraction of intelligence bearing planets where technology can develop; and $L$ = the mean lifetime of a technological civilization within the detection window.

An additional problem of the Drake equation is the time-independence of its terms (Cirkovic, 2004), which for example excludes the effects of the physico-chemical history of the galaxy (Forgan, 2009). Indeed, Cirkovic (2004) shows that the main shortcoming of the Drake equation is its lack of temporal structure, i.e., it fails to take into account various evolutionary processes that form a prerequisite for anything quantified by *f* parameters and $n_e$. This Drake equation's drawback was mentioned earlier by Franck Drake but the discussion of systematic biases following such simplification was avoided (Drake and Sobel, 1991).

In particular, not only some difficulties arising from changing one or more parameters values in Eq. 1 with time, but also the Drake equation doesn't provide any error estimation about the measured quantity. To be short, a estimation of *N=5* with a standard deviation (henceforth SD) *SD(N)<<1* is radically different from an estimation of *N=10* with a standard error of *SD(N)=10*. Recently, Maccone (2010) derives the first statistical Drake equation by associate each parameters with a random variable and then, given some assumptions, apply the Theorem Central Limit. However, this important new result doesn't take into account the temporal aspect of the processes of civilizations appearance. Here, we propose a first treatment of these evolutionary aspects by constructing a simple stochastic process which will be able to provide both a temporal structure to the Drake equation (i.e. introduce time in the Drake formula in order to obtain something like $N(t) = (R^* f_p n_e f_l f_i f_c L)(t)$ ) and a first standard error on *N(t)*.

## 2. A Stochastic Process Approach of the Drake Equation

### 2.1 Grouping the Drake parameters

When looking at the Drake equation given by equation (1), it is obvious that a kind of Bayesian structure underlying its construction (Shklovskii and Sagan, 1966). While the Bayesian structure of the SETI equation has been extensively described by Wilson (1984), the Drake equation has not been analyzed in this way. To begin with a heuristics approach, let us consider the three terms of the product $f_l f_i f_c$. For instance, $f_c$ is the estimate (because it is a frequency) of the probability that a technology arise on a planet, knowing the fact that intelligence has appear. Without worrying with formalism, it is something like $f_c = P(\text{technology}|\text{intelligence})$, where $P(A|B)$ is the conditional probability measure of the event A given B (Capiński and Kopp, 2002). In a similar way, $f_i$ is the estimate of the probability that intelligence arise on a planet, knowing the fact that life has appear ($f_i = P(\text{intelligence}|\text{life})$) and $f_l$ is the estimate of the probability that life arise on a planet, knowing the fact that we considered only potentially habitable planet ($f_l = P(\text{life}|\text{potentially habitable planet})$). More rigorously, if we consider the three set of events (i) $E_T$ (planet bears technology), (ii) $E_I$, (planet bears intelligence) and (iii) $E_L$ (planet bears life), all of them subset of the sample space of all potentially habitable planet, then it is straightforward that we have $E_T \subset E_I \subset E_L$. As a consequence of this underlying conditional structure, the product of this three previous terms is simply:

$$f_c f_i f_l = P(\text{technology}|\text{intelligence})P(\text{intelligence}|\text{life})P(\text{life}|\text{potentially habitable planet})$$
$$= P(\text{technology}|\text{potentially habitable planet})$$

This expression is an estimate of the probability that a technological civilization develop on a potentially habitable planet. All these preliminary remarks and heuristics approaches suggest that equation (1) parameters can be grouped together into two new parameters for which the meaning is straightforward:

1. $A = R^* f_p$ is the number of new planetary systems produced in the galaxy per year.

2. $B = n_e f_l f_i f_c$ is the number of advanced intelligent civilizations (henceforth AIC) that are able to communicate (and for which we can detect their communication) per planetary systems. AIC can be interpreted as instantiations of the sixth Dick's megatrajectory (Dick, 2003), in the same way Cirkovic's ATC are instantiations of the seventh's one (Cirkovic and Bradbury, 2006).

In other words, $AB$ is the number of new AIC produced per year. L is the average AIC lifetime. Historically, the Drake equation was rather written $N^* = Rf_s L$, where $R$ is the average rate of life-supportable star production, $f_s$ is the number of civilizations per suitable star and $L$ is still the average lifetime of an AIC. Equation (1) was established by Shklovskii and Sagan (1966) by expanding $f_s$.

AIC appearance occurs in a space which is by definition the Galactic Habitable Zone (henceforth GHZ). The concept of Galactic Habitable Zone was introduced a few years ago as an extension of the much older concept of Circumstellar Habitable Zone (Lineweaver et al., 2004). This location is usually considered to be an annulus, with inner radius 7 kpc and outer radius 9 kpc (1 kpc = 1000 pc ≈ 3000 light years) (Lineweaver et al., 2004; Forgan, 2011). However, other authors are pointing out that the physical processes underlying the former concept are hard to identify and that the entire Milky Way disk may well be suitable for complex life (Prantzos, 2008; Gowanlock, 2011). In this paper, we will consider the entire Milky Way disk to be suitable for the complex life, i.e. to be the GHZ. This will also allow us to use estimation of $R^*$ for the entire galaxy (Diehl et al., 2006).

This above new parameter grouping, the need for a temporal structure to the Drake equation and the reasonable assumption that AIC appearance should be roughly random in time and in space (this assumptions are discussed in the following part) strongly suggest that an AIC appearance mathematical model could be made by using a stochastic process like a Poisson process $\{N(t): t \geq 0\}$ with rate parameter $\lambda = AB$.

## 2.2 Poisson process

A Poisson process is a continuous-time stochastic process in which events occur continuously and independently of each others. Examples that are well-modeled as Poisson processes include the radioactive decay of atoms (Foata, 2002), Turing machine rules mutations (Glade *et al.*, 2009), the arrival of customers in a queue *n* telephone calls arriving at a switchboard and proteins evolution (Bastien, 2008; Ortet and Bastien, 2010). The Poisson process is a collection $\{N(t): t \geq 0\}$ of random variables, where $N(t)$ is the number of events, often called "top", that have occurred up to time *t* (starting from time *0*). The number of events between time *a* and time *b* is given as *N(b) – N(a)* and has a Poisson distribution. Each realization of the process $\{N(t): t \geq 0\}$ is a non-negative integer-valued step function that is non-decreasing in time. In our case, each "top" could be an AIC appearance. Hence, $N(t)$ would be the number of AIC that has appeared up to time *t*.

**Definition of a Homogeneous Poisson process**

The homogeneous Poisson process is one of the most well-known Lévy processes (Itô, 2004). A continuous-time counting process $\{N(t): t \geq 0\}$ will be called a Poisson process if it possesses the following properties:

1. $N(0) = 0$.

2. Independent increments (the numbers of occurrences counted in disjoint intervals are independent from each other).

3. Stationary increments (the probability distribution of the number of occurrences counted in any time interval only depends on the length of the interval).

4. No counted occurrences are simultaneous. More precisely, the process is locally continuous in probability, i.e., for all $t \geq 0$, $\lim_{h \to 0}\{P(N(t+h) - N(t)\} = 0$.

In our model, condition 1 means that we must begin the AIC count at a time when no previous AIC exists. Condition 2 means that each AIC evolves independently from each others. Condition 3 means that the number of AIC in a time interval does not depend on the date at which we sample this interval, that is to say invariance of physical law and global homogeneity of the space-time in the considered galaxy region, i.e. the Galactic Habitable Zone (Cirkovic, 2004; Gonzales *et al.*, 2001; Gowanlock, 2011; Lineweaver *et al.*, 2004; Prantzos, 2008; Vukotic and Cirkovic, 2007). If these conditions are satisfied, then we can deduce the following results.

**Consequences of this definition include:**

1. The probability distribution of *N*(*t*) is a Poisson distribution. That is to say $P(N(t+\tau) - N(t) = k) = \dfrac{(\lambda \tau)^k e^{-\lambda \tau}}{k!}$ where $N(t+\tau) - N(t)$ is the number of events between the time interval $]t, t+\tau]$ and $\lambda$ is the stochastic process rate parameter also called density or intensity). The product $\lambda \tau$ is called the parameter of the Poisson distribution.

2. The probability distribution of the waiting time until the next event occurrence is an exponential distribution.

3. The occurrences are distributed uniformly on any interval of time. (Note that N(*t*), the total number of occurrences, has a Poisson distribution over [0, *t*], whereas the location of an individual occurrence on *t* in [*a*,*b*] is uniform).

**Homogeneous and non-homogeneous Poisson process**

As recalled above, a homogeneous Poisson process is characterized by its rate parameter $\lambda$, which is the expected number of events (also called arrivals) that occur per unit time. Nevertheless, in general the rate parameter may change over time; such a process is called a non-homogeneous Poisson process or inhomogeneous Poisson process. In this case, the generalized rate function is given as $\lambda(t)$, where $\lambda(t)$ is a real continuous function of time (and hence, define on the positive part of the real axis). In this case, the above definition of a Poisson process remains unchanged except for the third condition (Stationary increments). Then, the three Poisson process conditions are:

1. $N(0) = 0$.

2. Independent increments (the numbers of occurrences counted in disjoint intervals are independent from each other).

3. Let $\rho(t) = \int_0^t \lambda(u)du$; Then for all pair *(s,t)* with $0 \leq s < t < +\infty$, the number $N(t) - N(s)$ of events occurring in $]s,t]$ is a Poisson random variable with parameter $\mu(]s,t]) = \rho(t) - \rho(s) = \int_s^t \lambda(u)du$.

This last condition imply the fourth of the definition of the homogeneous processes, that is to say: the process is locally continuous in probability.

Moreover, it can be demonstrate (Foata, 2002) that this condition is similar to the following:

*Condition 3bis.* for $h \to 0$, we have $P(N(t+h) - N(t) = 1) = \lambda(t) + o(h)$ and $P(N(t+h) - N(t) \geq 2) = o(h)$.

Usually, *m(t)* is called the renewable function. A remarkable result is that all non-homogeneous Poisson process can be transformed into a homogeneous Poisson process by a time transformation (Foata, 2002). Of course, a homogeneous Poisson process may be viewed as a special case when λ(t) = λ, a constant rate.

**2.3 AIC Birth and death process**

Here, we construct a simple stochastic process which will represents the stochastic appearance of advanced intelligent civilizations in the Milky Way as a function of time. More exactly, we will construct a model which will give us the number $C(t)$ of existing AIC for a given time $t$. A first reasonable hypothesis for this model is that there exist a time $t_0$ for which no AIC is present in the galaxy, i.e. $C(t_0) = 0$. So, without loss of generality, we can consider a stochastic process for which the first condition, $C(0) = 0$, for the process to be a Poisson process is true. A second hypothesis for this model is that AIC appearances (i.e. births) are independent from each other's (communication between them doesn't influence their birth nor their lifetime). As a consequence, a possible limitation of the present model could come from the fact that there is a legitimate case to be made that AIC numbers may violate this condition of the Poisson distribution: the longevity, may be significantly affected by discovery of a long-lived intelligent community, possibly leading to clustering in time. A third hypothesis is that the number of AIC in a time interval does not depend (at least locally) on the date at which we sample this interval.

With these three hypotheses, we can consider the stationary increments Poisson stochastic process $\{N(t): t \geq 0\}$ with density $\lambda > 0$ (Itô, 2004) where each "top" of the Poisson process is an AIC birth event. As the rate parameter $\lambda$ represents expected number of events that occur per unit time, it is clear that it is equal to the number of new AIC produced per year, that is to say the product *AB* of the upper new Drake equation parameter grouping. We also suppose that each AIC lifetime is a random variable *X*, i.e. all AIC during time are independents, identically distributed and are independent of the process $\{N(t): t \geq 0\}$. Let $S_k$ be the birth date of the $k^{th}$ AIC and $X_k$ its lifetime. Then, its death date is $S_k + X_k$. Let $S_0 := X_0 := 0$ and $C(t)$ be the number of AIC present at the time *t*. The question is to evaluate the probability law of $C(t)$, with the hypothesis $N(0) = 0$ (and hence $C(0) = 0$). The following theorem for the current AIC number can then be formulated (for the proof, see Appendix):

*<u>Theorem 1. Theorem for the current AIC number:</u> With the previous hypotheses and notations, the number $C(t)$ of AIC present at the time t is a Poisson distributed random variable with parameter $m(t) = \lambda \int_0^t r(u) du$, where $\lambda$ is the rate parameter of the AIC appearance Poisson stochastic process and $r(u)$ is the survival function of AIC lifetime random variable X.*

To resume, the Poisson stochastic process $\{N(t): t \geq 0\}$ which represents the appearance of new Advanced Intelligent Civilization in the Milky Way is combined with a random variable *X* which represents the lifetime of these new civilizations. The main results is that the number *C(t)* of AIC present at time t is a Poisson distributed random variable with parameter *m(t)*. This general result can be link to the classical Drake equation by the following. As t gets to infinity and using a classical theorem of probability about survival function (which stated that $\int_0^{+\infty} r(u)du = E[X]$; Skorokhod and Prokhorov, 2004), we observe that $\lim_{t \to +\infty} m(t) = \lambda \int_0^{+\infty} r(u)du = \lambda E[X]$. As a consequence, $C(t)$ is going toward a Poisson random variable *V* with parameter $\lambda E[X]$. This result is exactly the Drake equation with $E[V] = \lim_{t \to +\infty} m(t)$, $E[X] = L$ and $\lambda = AB$. Interestingly, if one assume that the lifetime of any galactic civilization is finite, that is to say has an upper bound. Then, it exist a number $t_M$ for which $r(t_M)=0$ and so the limit value of m(t) will be reached at finite time, that is to say we will have $\int_0^{t_M} r(u)du = E[X]$.

## 3. Discussion

**Mean and variance of the number of AIC**

As stated above, as *t* becomes larger than the AIC maximum lifetime, the previous approximation becomes exact. So we have $E[V(t)] = VAR[V(t)] = \lambda E[X]$ (the mean and the variance of a Poisson distributed random variable are equal;

Skorokhod and Prokhorov, 2004) and the three terms of the equality are time independent. As a consequence, we can study the coefficient of variation $\varepsilon$ of the $C(t)$ (also named fluctuation around the mean) which is the ratio between the standard deviation and the mean of the stochastic process. Here, we have

$$\varepsilon = \frac{1}{\sqrt{\lambda E[X]}} = \frac{1}{\sqrt{R^* f_p n_e f_l f_i f_c E[X]}}$$

With the 1961's Drake parameters estimation (see table 1), $\lambda = 0.001$ and hence, depending on the AIC lifetime, we can have

1. with $E[X] = 200$, we get $E[V(t)] = 0.2$ and $\varepsilon = 2.23$,

2. with $E[X] = 10000$, we get $E[V(t)] = 10$ and $\varepsilon = 0.31$

A error of magnitude one order in any parameter can lead to a estimation of $E[V(t)]$ and $\varepsilon$ equal to 1. For example, a lower bound can be estimated for $E[X]$ while considering the span time between now and the invention of the parabolic telescope, *i.e.* radioastronomy (Reber and Conklin, 1938). This gives $E[X] = 73$, and so $E[V(t)] = 0.072$ and $\varepsilon = 3.9$.

The dramatic effect of parameter evolution estimations on possible value of $N(t)$ can be seen in the following. Indeed, with more recent estimations (Diehl *et al.*, 2006; Maccone, 2010), we get $\lambda = 0.07$ and hence, we can have

1. with $E[X] = 200$, we get $E[V(t)] = 14$ and $\varepsilon = 0.27$,

2. with $E[X] = 10000$, we get $E[V(t)] = 700$ and $\varepsilon = 0.04$

3. with $E[X] = 73$, we get $E[V(t)] = 5$ and $\varepsilon = 0.45$

All these considerations give a large probability for detection of another AIC, depending on the reliability of the new estimations. Starting from $V(t_0) = 0$, the Poisson stochastic process theory allow us to estimate the average mean time for the occurrence of a new AIC appearance, which is given by the inverse of $\lambda$ (Foata, 2002). With the 1961's Drake parameters estimation (resp. recent estimation), this average time is equal to should be 1,000 years (resp. 14 years). Apply to the SETI research program, if we suppose that we are alone in our universe, the average mean time for the occurrence of an other AIC should be roughly of 14 years with recent parameter estimations (resp. 1,000 years for the 1961's Drake estimation), with a standard error of the same order error.

## 4. Conclusion

The proposed model allows a first analytic estimation of the standard deviation of the number of Galactic civilization estimate. In addition, it provides a temporal structure of the Drake equation which can help the studying of the influence of several effects on the number of Galactic civilization estimate. A important case is the notion of global regulation mechanism (i.e. a dynamical process preventing uniform emergence and development of life all over the Galaxy; Annis, 1999; Vukotic and Cirkovic, 2008). Vukotic and Cirkovic (2007) investigated the effects of a particular global regulation mechanism, the Galactic gamma-ray bursts (GRBs) (colossal explosions caused either by terminal collapse of supermassive objects or mergers of binary neutron stars), on the temporal distribution of hypothetical inhabited planets, using simple Monte Carlo numerical experiments. Here, GRB is clearly just one of the possible physical processes for resetting astrobiological clocks. They obtain that the times required for biological evolution on habitable planets of the Milky Way are highly correlated. More precisely, using simulations cosmological observations (Bromm and Loeb, 2002), they demonstrated that the correlation (and so the covariance $\text{cov}(t_b, t_*)$) between the biological timescale $t_b$ and the astrophysical timescale $t_*$ is non zero. Using the distribution of GRB over the time, a analytic approach would be to compute the random time $T$ since the last GRB event. Assuming that T is independent from both $\{N(t) : t \geq 0\}$, the Poisson stochastic process which represents the appearance of new AIC, and $X$ which represents the lifetime of these new civilizations, the theorem 1 can be replaced by the more general form:

*Theorem 2. Theorem for the current AIC number under Global Regulation Mechanisms: With the previous hypotheses and notations, the number $C(T)$ of AIC present at the time T is a Poisson distributed random variable with parameter $m(T) = \lambda \int_0^T r(u) du$, where $\lambda$ is the rate parameter of the AIC appearance Poisson stochastic process, $r(u)$ is the survival function of AIC lifetime random variable X and T is the random variable 'time elapsed since the last major global regulation event'.*

This result follow from the above fact that $E[C(t)] = m(t) = \lambda \int_0^t r(u) du$, which can be rewrite $E[C(T)|T=t] = m(t)$, where E[.|.] is the conditional expectation operator. This last formula is called the conditional expectation of $C$ given $T = t$. Since we don't know the true value of $t$ since the last GRB event, we must consider the new random variable $m \circ T = m(T) = E[C|T]$. Future works will explore this result. Especially, it would be suitable to obtained formula on the

form like $E[C(T)] = \lambda E[\int_0^T r(u)du]$ where the expectation is computed using the global regulation event (in particular GRB events) distribution over the time. This analytic approach could provide an accurate analysis of classical Monte-Carlo simulations (Vukotic and Cirkovic, 2008; Forgan, 2009; Forgan, 2011; Hair, 2011). For instance, Hair (2011) and Forgan (2011) had proposed in their two models that the distribution of the civilization arrival times is Gaussian-distributed. In technique, this is equivalent to allow the AIC appearance rate parameter $\lambda$, to varying in time. More precisely, the rate parameter $\lambda(t)$ corresponding to the Forgan model (2011) should be a increasing function on $[0,\mu]$, where $\mu$ is the mean of the Hair (2011) and Forgan (2011) arrival time Gaussian distribution ($\mu$ has the same order of magnitude than the Hubble Time, $t_H$= 13,700Myr), and a decreasing function on $]\mu,+\infty[$. Nevertheless, this work is mainly a first approach to model AIC appearance but future studies would have to address the fact that AIC appearance on habitable planets should be correlated with the Galaxy's star formation history (Heavens et al., 2004; Juneau et al., 2005; Vukotic and Cirkovic, 2007) and the location of the Galactic Habitable Zone (Gonzales et al., 2001; Gowanlock, 2011; Lineweaver et al., 2004; Prantzos, 2008;). For instance, Planet formation and star formation could be included in this first AIC appearance model by extracting the original Drake parameter $R^*$, $f_p$ and $n_e$ from $\lambda$ and let them varying time.

## 5. Appendix: demonstration of the theorem

The proof of the theorem can be established in three parts (Foata, 2002).

**Lemma:** *Let $r(u) := P\{X \geq u\}$ be the survival function of X and $\bar{r}(t) := \frac{1}{t}\int_0^t r(u)du$.*

*If U is a uniform random variable on $[0,t]$, independent of X, the survival function of U+X is given by: $P\{U + X \geq t\} = \bar{r}(t)$.*

<u>Proof:</u> Following the fact that the density of U is $\frac{1}{t}I_{[0,t[}$, where $I_{[0,t[}(x)$ is 0 outside the interval [0,t[, we have

$$P\{U + X \geq t\} = \frac{1}{t}\int_0^t P\{U + X \geq t | U = s\}ds = \frac{1}{t}\int_0^t P\{X \geq t - s | U = s\}ds = \frac{1}{t}\int_0^t P\{X \geq t - s\}ds$$

$$P\{U + X \geq t\} = \frac{1}{t}\int_0^t r(t-s)ds = \frac{1}{t}\int_0^t r(u)du \text{ which is equal to } \bar{r}(t) \text{ by definition.}$$

**Lemma:** *The generating function $h(u) := E[u^{C(t)}]$ of $C(t)$ is given by $e^{-\lambda t \bar{r}(t)(1-u)}$.*

<u>Proof:</u> For $k \geq 0$, let $Y_k := I_{\{S_k + X_k \geq t\}}$. Obviously we have $C(t) = \sum_{k=0}^{N(t)} Y_k$, and so

$$h(u) = E[u^{C(t)}] = \sum_{n \geq 0} E[u^{C(t)} | N(t) = n]P\{N(t) = n\} = \sum_{n \geq 0} E[u^{Y_0 + Y_1 + \ldots + Y_n} | N(t) = n]P\{N(t) = n\}$$

$$h(u) = \sum_{n \geq 0} E\left[\prod_{k=0}^{n} u^{Y_k} \bigg| N(t) = n\right] P\{N(t) = n\}.$$

Conditionally to the event $\{N(t) = n\}$ ($n \geq 1$), the system $(S_1, S_2, \ldots, S_n)$ has the same distribution than the system $(U_1, U_2, \ldots, U_n)$ of independent and uniformly distributed on [0,t] random variables.

For $1 \leq k \leq n$, let $Z_k := I_{\{U_k + X_k \geq t\}}$. We have

$$h(u) = \sum_{n \geq 0} E\left[\prod_{k=0}^{n} u^{Z_k}\right] P\{N(t) = n\} = \sum_{n \geq 0} (E[u^{Z_1}])^n P\{N(t) = n\}.$$

But $Z_1$ is a Bernoulli random variable with parameter $P\{U_1 + X_1 \geq t\} = \bar{r}(t)$. Hence, $E[u^{Z_1}]$ is the generating function of the $Z_1$ random variable and is given by

$$E[u^{Z_1}] = \sum_{k=0}^{\infty} P(z=k) z_1^k = 1 - \bar{r}(t) + u.\bar{r}(t)$$ (Koroliouk, 1978). Finally, we get

$$h(u) = \sum_{n \geq 0} \frac{e^{-\lambda t}(\lambda t)^n}{n!}(1 - \bar{r}(t) + u.\bar{r}(t))^n = e^{-\lambda t \bar{r}(t)(1-u)}$$

<u>Proof of the theorem:</u> We can write $h(u) = e^{-m(t)(1-u)}$ which is the generating function of a Composed Poisson distribution (Koroliouk, 1978) and so, is a Poisson distributed random variable with parameter $m(t) = \lambda \int_0^t r(u) du$.

**Acknowledgements**

OB was supported by the french Agence Nationale de la Recherche, as part of the ReGal project.

**Author Disclosure Statement**

No competing financial interests exist.

**Table 1.** Drake Equation Parameter Estimations. References: 1. Drake and Sobel (1991) 2. Diehl et al. (2006). 3. Maccone (2010)

| Parameters | Significations | 1961's estimation [1] | Recent estimations |
|---|---|---|---|
| N | the number of Galactic civilizations who can communicate with Earth | | |
| $R^*$ | the average rate of star formation per year in our galaxy | 10 per year | 7 per year [2] |
| $f_p$ | the fraction of stars that host planetary systems | 0.5 | 0.5 [3] |
| $n_e$ | the number of planets in each system that are potentially habitable | 2 | 1 [3] |
| $f_l$ | the fraction of habitable planets where life originates and becomes complex | 1 | 0.5 [3] |
| $f_i$ | the fraction of life-bearing planets that bear intelligence | 0.01 | 0.2 [3] |
| $f_c$ | the fraction of intelligence bearing planets where technology can develop | 0.01 | 0.2 [3] |
| L | the mean lifetime of a technological civilization within the detection window | 10,000 | See text |